\def\year{2021}\relax
\documentclass[letterpaper]{article} % DO NOT CHANGE THIS
\usepackage{aaai21}  % DO NOT CHANGE THIS
\usepackage{times}  % DO NOT CHANGE THIS
\usepackage{helvet} % DO NOT CHANGE THIS
\usepackage{courier}  % DO NOT CHANGE THIS
\usepackage[hyphens]{url}  % DO NOT CHANGE THIS
\usepackage{graphicx} % DO NOT CHANGE THIS
\def\UrlFont{\rm}  % DO NOT CHANGE THIS
\usepackage{natbib}  % DO NOT CHANGE THIS AND DO NOT ADD ANY OPTIONS TO IT
\usepackage{caption} % DO NOT CHANGE THIS AND DO NOT ADD ANY OPTIONS TO IT
\usepackage{boldline,multirow}
\usepackage{ragged2e}
\usepackage{amsmath}
\usepackage{array}
\title{Qualitative Investigation in Explainable Artificial Intelligence: A Bit More Insight from Social Science}
\author{

    Adam J. Johs,
    Denise E. Agosto,
    Rosina O. Weber\\
    
}
\affiliations{
    Department of Information Science, College of Computing and Informatics, Drexel University, Philadelphia, USA\\

   \{ ajj37, dea22, rw37 \}@drexel.edu
}
\begin{document}

\maketitle

\begin{abstract}
We present a focused analysis of user studies in explainable artificial intelligence (XAI) entailing qualitative investigation. We draw on social science corpora to suggest ways for improving the rigor of studies where XAI researchers use observations, interviews, focus groups, and/or questionnaires to capture qualitative data. We contextualize the presentation of the XAI papers included in our analysis according to the components of rigor described in the qualitative research literature: 1) underlying theories or frameworks, 2) methodological approaches, 3) data collection methods, and 4) data analysis processes. The results of our analysis support calls from others in the XAI community advocating for collaboration with experts from social disciplines to bolster rigor and effectiveness in user studies.
\end{abstract}

\section{Introduction}\noindent One impetus for advancing the field of explainable artificial intelligence (XAI) has been the incorporation of insights from social science corpora into XAI research \cite{MillerHoweSonenberg2017,hoffman2017explaining,hoffman2017explaining2,degraaf2017people}. Cognitive scientists with expertise in cognitive psychology have underscored the nuance of how explanations can be used in XAI \cite{Byrne2019}. Teams of social scientists are serving as pillars to XAI research programs situated in national defense agencies \cite{HoffmanMuellerKleinLitman2018,MuellerHoffmanClanceyEmreyKlein2019,GunningAha2019}. Findings synthesized from social science literature now offer a foundation for researchers to inform investigation in XAI \cite{Miller2019}; experts have also speculated on XAI systems occupying key social capacities in the future \cite{GunningStefikChoiMillerStumpfYang2019}.

This paper extends the stream of XAI literature composed of papers from authors who have drawn from the social sciences. Our chief contribution is a focused analysis of the XAI literature on qualitative investigation. In a vein akin to \citet{MillerHoweSonenberg2017}, this work is not intended as an exhaustive literature review, but a focused analysis of papers to spotlight the current opportunities in XAI that can be capitalized on to enrich user studies. `User studies' is a broad term encompassing a gamut of evaluations, like experiments and field studies, where users are involved \cite{preece2015interaction}. We use `user studies' also in reference to studies conducted for purposes other than evaluation (\emph{e.g.}, exploratory investigation, for discovery, to collect requirements). Coupled with the knowledge drawn from the social science literature, we argue the findings presented in this paper are critical to catalyzing systematic qualitative investigation in XAI; our ultimate aim is to bolster the rigor and effectiveness of user studies in XAI.

Our motivation aligns with arguments raised by others in the XAI community regarding user studies \cite{Miller2019,HoffmanMuellerKleinLitman2018,PayrovnaziriChenRengifoMorenoMillerBianChenLiuHe2020,BhattAndrusWellerXiang2020}. Of note, \citet{PayrovnaziriChenRengifoMorenoMillerBianChenLiuHe2020} underline the importance for interdisciplinary cooperation (between AI researchers and medical professionals) toward maximizing the effectiveness of XAI methods in the context of medicine. In their description of a recent multinational, interdisciplinary workshop on explainable machine learning, \citet{BhattAndrusWellerXiang2020} punctuate the need for broad collaboration with experts from social disciplines if the field of XAI is to ensure rigor and effectiveness in user evaluations.

We marry the perspectives of \citet{PayrovnaziriChenRengifoMorenoMillerBianChenLiuHe2020} and \citet{BhattAndrusWellerXiang2020} with the correlative outlook of researchers not fixed in the traditional bounds of AI. As a human-computer interaction (HCI) expert, \citet{Xu2019} emphasizes the importance of spurring interdisciplinary collaboration in XAI to assure rigor in experiments involving users and validity in research protocols. We share the sentiments of \citet{PayrovnaziriChenRengifoMorenoMillerBianChenLiuHe2020}, \citet{BhattAndrusWellerXiang2020}, and \citet{Xu2019}, drawing attention to the qualitative research literature in hopes of galvanizing a systematic approach to qualitative investigation in XAI.

\section{Qualitative Research: Background, Data, and Methods}
\subsubsection{Background} There are numerous examples in the AI literature of researchers using ‘qualitative’ to validly describe analyses that extend beyond quantitative aspects \cite{LapuschkinBinderMontavonMullerKRSamek2016,Forbus1984,Druzdzel1996,HowellWoodwardChoueiryYu2018,ClarkMatwin1993,RizzoLongo2018,NguyenDosovitskiyYosinskiBroxClune2016,GilpinBauYuanBajwaSpecterKagal2018,ParkHendricksAkataRohrbachSchieleDarrellRohrbach2018,WuSunLiSongLi2017,KoulGreydanusFern2018,UstunRudin2016,KwonChoiKimChoiKimKwonSunChoo2018}. In qualitative research, however, ‘qualitative’ refers to a naturalistic orientation to the world used to study the human facets of a topic—namely, the meanings brought by research participants to the phenomena under investigation \cite{LincolnGuba1985,Given2008,AtkinsonDelamont2010}.

\subsubsection{Qualitative Data} Broadly distinguishing between quantitative data and qualitative data, the former is typically numerical, whereas the latter is typically nonnumerical \cite{Schreiber2008}. Many qualitative researchers also study phenomena quantitatively, for example, using statistical analyses to describe data, and as a means for determining consistency of findings across data collection methods ({\it e.g}., interviews vis-a-vis questionnaires) \cite{Donmoyer2008}. In qualitative research, many data analysis processes involve some form of ‘coding’—a fundamental data analysis technique researchers use to attribute meaning to data to support categorization, pattern identification, and/or theory development \cite{MilesHuberman,Saldana2013,Given2008}. Coding is a cyclical process encompassing 30+ coding methods researchers can select from to analyze qualitative data \cite{Saldana2013}; crucially, ‘code’ in qualitative data analysis is not to be conflated with ‘code’ as is used in semiotics, nor with ‘code’ as is used in computer programming.

\subsubsection{Qualitative Methods} As XAI studies may adopt a quantitative, qualitative, or mixed-methods design \cite{ChromikSchuessler2020}, we limit our focus to XAI papers where \emph{some} degree of qualitative design was adopted by the authors as part of user studies. In line with \citet{Mahoney1997}, we delimit the scope of our analysis (Section 3) to XAI papers where authors incorporated any form of the three most common methods used for qualitative investigation: 1) observations, 2) interviews, and/or 3) focus groups.

Observations are used in qualitative investigation to inform understanding of research context, affording researchers the opportunity to collect data on the processes, participant behaviors, or programs under study \cite{Mahoney1997}. Observations can be conducted in artificial ({\it e.g}., laboratory) settings, but generally occur in natural settings to capture data as data exist in the real-world \cite{McKechnie2008}. Observations afford researchers a means to learn about aspects that may not arise in interviews or focus groups due to the reticence and/or unawareness of participants \cite{Mahoney1997}.

Interviews are used to capture insights directly from research participants \cite{Mahoney1997}, conducted in-person, via telephone, or on the Web \cite{Brinkmann2008}. Interview formats can be structured (rigid), semi-structured (flexible), or unstructured (free-form) \cite{LazarFengHochheiser2017}. Structured interviews are useful for evaluating systems when a streamlined comparison of responses from multiple users is required; unstructured interviews are useful for obtaining deep insights and affording participants a chance to explore topics they (participants) deem significant; and semi-structured interviews offer utility to researchers who are unfamiliar with a domain or user group and are interested in obtaining deep insights from participants, absent the unbounded nature of unstructured interviews, and absent the limiting nature of structured interviews \cite{LazarFengHochheiser2017}. Most interviews in qualitative research are semi-structured, as the flexibility of semi-structured interviews afford a versatile happy medium (between the structured and unstructured formats) for data collection \cite{Brinkmann2008}. Interviewing offers researchers utility for informing various research activities including initial exploration and requirements gathering \cite{LazarFengHochheiser2017}.

Focus groups, sometimes referred to as ‘group interviewing’ \cite{Morgan2008}, employ aspects of qualitative interviewing to elicit (otherwise undiscoverable) data from participants by leveraging the nature of group dynamics \cite{Mahoney1997}. The social influences amongst focus group participants can yield insights that would not organically emerge in individual settings \cite{Morgan1997}. Thoughtful facilitation of focus groups is required not only to capture useful data, but to moderate conflict, stimulate engagement, sustain conversation, manage time, and navigate through interview guides \cite{LazarFengHochheiser2017}. Focus groups can be conducted virtually and generally encompass 4--10 participants, lasting no longer than 2 hours \cite{Liamputtong2011}.

There is some overlap between observations, interviews, focus groups and a fourth research method germane to XAI (as evidenced by \citeauthor{HoffmanMuellerKleinLitman2018} \citeyear{HoffmanMuellerKleinLitman2018}): questionnaires; our usage of ‘questionnaire’ is synonymous with the usage of ‘survey’ in the context of research methods (corresponding to the usage of \citeauthor{LazarFengHochheiser2017} \citeyear{LazarFengHochheiser2017}). Per \citet{LazarFengHochheiser2017}, the probabilistic vs. non-probabilistic structure, number of responses obtained, and inclusion of open-ended vs. close-ended questions determine 1) the type of data captured via questionnaires ({\it i.e}., quantitative data, qualitative data, or both quantitative data and qualitative data) and 2) how the questionnaire data are analyzed. Open-ended questions yield qualitative data and must be qualitatively coded as part of analysis \cite{LazarFengHochheiser2017}. 

Researchers often collect qualitative data within experimental designs toward understanding how participants experienced variables (\emph{e.g.}, interventions)---this is a form of mixed methods study design termed `embedded mixed methods design' \cite{Creswell2008}. We refer to the questionnaires disseminated or administered to respondents, and the tasks where participants textually answer open-ended questions pertaining to experimental variables, as `questionnaire tasks.' Accordingly, as part of our analysis (Section 3), we also include XAI papers where any variant of a questionnaire task was employed during a user study to capture qualitative data.

\section{Focused Analysis of Qualitative Investigation in XAI}

\begin{table*}[t!]
\small
\resizebox{\textwidth}{!}{%
\begin{tabular}{>{\raggedright\arraybackslash}p{0.275\textwidth}|>{\raggedright\arraybackslash}p{0.075\textwidth}|>{\raggedright\arraybackslash}p{0.2\textwidth}|>{\raggedright\arraybackslash}p{0.2\textwidth}|>{\raggedright\arraybackslash}p{0.15\textwidth}|>{\raggedright\arraybackslash}p{0.15\textwidth}>{\raggedright\arraybackslash}p{0.15\textwidth}}
\hlineB{4}
\textbf{PAPER} & CITED BY & UNDERLYING THEORY OR FRAMEWORK & METHODOLOGICAL APPROACH  & DATA COLLECTION METHOD & DATA ANALYSIS PROCESS \\\hlineB{4}

\textbf{\citet{AlshehriMillerVeredAlamri2020}} & \emph{Not in GS} & Folk Theory of Mind and Behavior \cite{TateChuck2006EtPP} & ―  & Questionnaire Task & Thematic Analysis \\ \hline

\textbf{\citet{ConatiBarralPutnamRieger2020}} & \emph{Not in GS} & Framework for User Modeling and Adaptation \cite{kardan2015providing} & ― & Questionnaire Task & ― \\ \hline

\textbf{\citet{doyle2006evaluation}} (\emph{Evaluation Part 1, Evaluation Part 2, and Counter Example Evaluation}) & 46 & Explanation utility framework \cite{doyle2004explanation} & ― & Questionnaire Task & ― \\ \hline

\textbf{\citet{DzindoletPierceBeckDawe2002}} (\emph{Study 2}) & 256 & Framework to predict automation use \cite{dzindolet2001framework} & ― & Questionnaire Task & Content Analysis \\ \hline

\textbf{\citet{DzindoletPetersonPomrankyPierceBeck2003}} (\emph{Study 2}) & 840 & ― & ― & Questionnaire Task & Content Analysis \\ \hline

\textbf{\citet{eisenstadt2018explainable}} & 7 & Explanation patterns framework \cite{cassens2007designing} & ― & Questionnaire Task & ― \\ \hline

\textbf{\citet{HallHarborneTomsettGaleticQuintanaAmateNottlePreece2019}} & 10 & Conceptual framework characterizing explainability \cite{bohlender2019towards} & ― & Semi-Structured Interviews & ― \\ \hline

\textbf{\citet{HerlockerKonstanRiedl2000}} (\emph{Experiment 2}) & 1919 & ― & ― & Questionnaire Task & ― \\ \hline

\textbf{\citet{HuberWeitzAndreAmir2020}} & \emph{Not in GS} & ― & ― & Questionnaire Task & Content Analysis (Summative) \\ \hline

\textbf{\citet{KimGlassmanJohnsonShah2015}} (\emph{Pilot Study}) & 19 & ― & ― & Questionnaire Task & ― \\ \hline

\textbf{\citet{KimShahDoshiVelez2015}} (\emph{Pilot Study}) & 78 & ― & ― & Questionnaire Task & ― \\ \hline

\textbf{\citet{KrausePererNg2016}} & 205 & ― & Case Study & Interviews (\emph{*format unspecified}) & Exploratory Data Analysis \\ \hline

\textbf{\citet{kulesza2013too}} & 147 & ― & Grounded Theory & Questionnaire Task & Coding (\emph{*method(s) unspecified}) \\ \hline

\textbf{\citet{Lakkaraju2020}} (\emph{Human Evaluation of Trust in Black Box}) & 24 & Model Understanding through Subspace Explanations (MUSE) framework \cite{Lakkaraju2019} & ― & Questionnaire Task & ― \\ \hline

\textbf{\citet{pacer2013evaluating}} (\emph{Experiment 1}) & 33 & Various explanation models associated with probability theory, notably, \citet{Pearl1988} & ― & Questionnaire Task & Coding (\emph{*method(s) unspecified}) \\ \hline

\textbf{\citet{PutnamConati2019}} & 16 & User modeling framework \cite{kardan2015providing} & ― & Questionnaire Task & ― \\ \hline

\textbf{\citet{PutnamConati2019}} & 16 & User modeling framework \cite{kardan2015providing} & ― & Interviews (\emph{*format unspecified}) & ― \\ \hline

\textbf{\citet{ribeiro2016should}} (\emph{Husky vs. Wolf task}) & 4326 & ― & ― & Questionnaire Task & ― \\ \hline

\textbf{\citet{SilvaLelisBowling2020}} & \emph{Not in GS} & ― & ― & Questionnaire Task & ― \\ \hline

\textbf{\citet{singh2020quantitative}} & 1 & ― & ― & Questionnaire Task & ― \\ \hline

\textbf{\citet{TonekaboniJoshiMcCraddenGoldenberg2019}} & 40 & ― & Mixed Methods \cite{leech2009typology} & Interviews (\emph{*format unspecified}) & ― \\ \hline

\textbf{\citet{TullioDeyChaleckiFogarty2007}} & 134 & Theory of developing mental models for expert users \cite{moray1987intelligent} & Field Research & Semi-Structured Interviews & Coding (\emph{*method(s) unspecified}) \\ \hline

\textbf{\citet{williams2016axis}} & 104 & Learnersourcing framework \cite{kim2015learnersourcing} & ― & Questionnaire Task & ― \\ \hline

\textbf{\citet{williams2016axis}} & 104 & Learnersourcing framework \cite{kim2015learnersourcing} & ― & Semi-Structured Interview & ― \\ \hline

\end{tabular}%
}
\label{tab2}
\caption{A depiction of the papers in the XAI literature where some form of qualitative investigation was employed as part of a user study, along with the corresponding components of qualitative research rigor discussed in each paper}
\end{table*}

Myriad criteria and perspectives have been put forward to guide how quality in qualitative research should be approached \cite{Guba1981,Given2008,AtkinsonDelamont2010,Flick2007,Tracy2010,Flick2018,MaysPope2000,DingwallMurphyWatsonGreatbatchParker1998,Seale1999,Patton2003}. Though a synthesis of the different viewpoints on qualitative research quality is out of scope in this paper, we leverage the discussions of rigor as a fundamental criterion of qualitative research quality to contextualize presentation of the XAI papers included in our analysis.

Rigor as a criterion for qualitative research quality can be characterized as the clear description of steps in one’s research, including, appropriateness of the selected data collection method \cite{SaumureGiven2008}; \citet{Tracy2010} cites the proper utilization of theoretical constructs, data, samples, contexts, and data analysis processes as components of rigor. Hence, we conducted our analysis to illustrate the underlying theories or frameworks, qualitative data collection methods, and qualitative data analysis processes used thus far by XAI researchers as part of user studies. And because it is encouraged to use best practices of qualitative methodology as a language to dialogue with quantitatively-grounded research communities \cite{Tracy2010}, we also include qualitative methodological approaches in our analysis.

To begin gathering what has been contributed to the XAI literature on qualitative investigation, we employed a two-pronged approach: 1) examining the findings of recent XAI reviews and 2) exploring the papers published in XAI venues. The reviews examined were \citet{Miller2019,MuellerHoffmanClanceyEmreyKlein2019,ViloneLongo2020,PayrovnaziriChenRengifoMorenoMillerBianChenLiuHe2020,ArrietaDiazRodriguezNSerBennetotTabikBarbadoGarciaGilLopezMolinaBenjaminsChatila,AdadiBerrada2018,GuidottiMonrealeRuggieriTuriniGiannottiPedreschi2018,Lipton2018,KeaneKenny2019,BiranCotton2017,ChakrabortyTomsettRaghavendraHarborneAlzantotCeruttiSrivastavaPreeceJulierRaoKelleyBrainesSensoyWillisGurram2017,DoshiVelezKim2017,GilpinBauYuanBajwaSpecterKagal2018,AlonsoCastielloMencar2018,AnjomshoaeNajjarCalvaresiFramlingK2019,AbdulVermeulenWangLimKankanhalli2018,RiberaLapedriza2019,DoilovicFKBrcicMHlupicN2018,IsraelsenAhmed2019, SormoFCassensAamodt2005,RothBerghofer2004,TjoaGuan2019,PuiuttaVeith2020,ClinciuHastie2019,LongoGoebelLecueKiesebergHolzinger2020,Mathews2019,ZhangZhu2018,DoranSchulzBesold2017,XuUszkoreitDuFanZhaoZhu2019,CuiLeeHsieh2019}; and \citet{TintarevMasthoff2007}.

The papers explored were from (including but not limited to) the International Joint Conference on Artificial Intelligence (IJCAI); Association for the Advancement of Artificial Intelligence (AAAI) Conference on AI; ACM Conference on Intelligent User Interfaces (IUI); International Conference on Case-Based Reasoning (ICCBR); International Conference on Autonomous Agents and Multiagent Systems (AAMAS); Conference on Computer Vision and Pattern Recognition (CVPR); Conference on Neural Information Processing Systems (NIPS); and Association for Computing Machinery (ACM)/AAAI conference on Artificial Intelligence, Ethics, and Society (AIES). We prioritized our analysis on AI and computer science venues.

Any reference of qualitative investigation in the XAI papers found via the aforementioned reviews or venues were examined for inclusion in our analysis (as per the contextualization described at the beginning of this section). We also submitted various search queries to Google Scholar (GS), our university libraries, and the DuckDuckGo Internet search engine to expand the scope of our analysis (for example, \emph{qualitative investigation AND explainable AI}, \emph{qualitative investigation AND XAI}, \emph{qualitative explainable artificial intelligence}, \emph{qualitative evaluation AND explainable artificial intelligence}, \emph{qualitative evaluation AND explainable AI}, \emph{user evaluation AND explainable AI}, \emph{explainable AI user evaluation}, \emph{qualitative AND XAI}). \citet{Given2008} was referenced to guide our determination of whether a methodological approach, data collection method, or data analysis process used by the authors in any of the retrieved papers warranted inclusion in our analysis. The human-centered XAI papers compiled by \citet{MuellerHoffmanClanceyEmreyKlein2019} and \citet{ViloneLongo2020} served as valuable bases for our efforts.

Table 1 depicts the results of our analysis. Dashes are used to denote the components of qualitative research rigor not discussed in the reviewed papers. Only underlying theories or frameworks used to inform the design of user studies were included. Duplicate rows were created for papers entailing more than one qualitative data collection method (as described in Section 2). And for any papers detailing more than one study, the specific study where qualitative investigation was employed is included next to the citation in the corresponding row. To afford the reader a sense of how much attention has been directed to the papers included in our analysis, we also inserted a column of citation (cited by) counts found in GS; citation counts are current as of Dec. 14, 2020, and not every paper listed had a record in GS at the time of compilation (denoted with \emph{`Not in GS'} in Table 1).

\section{Discussion}
\subsubsection{Findings} The most common data collection method used by XAI researchers in the papers reviewed was instantiations of the questionnaire task. The underlying theories and frameworks used to inform the design of user studies spanned various domains---{\it e.g}., psychology, AI, human-centered computing. Many authors did not specify the format of interviews and/or method(s) of coding used---key details that influence interpretation of results from qualitative investigation. We did not see use of observations or focus groups (group interviewing) for data collection in any of the XAI papers included in our analysis.

As the papers in our analysis were assessed for inclusion based on whether each of the following elements were satisfied:
\begin{enumerate}
  \item Some degree of \emph{qualitative study design} was adopted by the authors
  \item \emph{Studies with users} were conducted
  \item \emph{Observations, interviews, focus groups, or questionnaires} were used to capture qualitative data
\end{enumerate}
notable papers were excluded from our analysis and the depiction in Table 1. The work of \citet{MadumalMillerSonenbergVetere2019} was excluded because grounded theory and coding were used to derive a dialogue framework from existing data---not as part of a user study---and the data yielded from the Wizard of Oz experiment \cite{Dahlback1993} were analyzed quantitatively.

\citet{kulesza2011oriented} used a think aloud protocol to conduct a study where participants debugged an intelligent assistant. \citet{bunt2012explanations} used semi-structured interviews and diaries, informed by a contextual inquiry approach, to investigate the explanations of low-cost intelligent interactive systems (IIS). \emph{Diaries} \cite{LazarFengHochheiser2017} and the \emph{think aloud protocol} \cite{ericsson1984protocol} are additional methods for capturing qualitative data beyond the scope of this paper, as is the design approach of contextual inquiry \cite{holtzblatt1997contextual}.

\citet{ZhouShengHowley2020} leveraged the user-centered design process as part of designing an XAI system. \citet{kulesza2010explanatory} and \citet{shinsel2011mini} both used programming-oriented approaches to inform qualitative investigation. The \emph{user-centered design} process \cite{abras2004user} and computer programming methodologies also extend beyond the qualitative research corpora we are drawing from in this paper. 

\subsubsection{Recommendations for the XAI Community} The results of our analysis underline the necessity for improving the rigor of qualitative investigation employed in XAI as part of user studies. Several of the papers analyzed lacked details required to assess qualitative research rigor. When presenting interview-based studies, investigators must specify the type of interviews conducted (structured, semi-structured, unstructured, or focus group interviewing); an overview of the interview guides used, including the main interview questions asked; and clear representations of the data collected, including data that supported major findings, and data that contradicted major findings.

Many of the papers in our analysis also lacked details describing the methodological approaches and data analysis processes employed. For example, most qualitative researchers across disciplines use thematic analysis and inductive coding as data analysis processes for qualitative investigation. Importantly, `inductive coding' only describes the process for analyzing qualitative data, and, in and of itself, thematic analysis does not provide a theoretical basis for analysis. `Thematic analysis' also does not explain how themes are identified, how themes evolve as coding progresses, or connections between the research questions and coding approaches used. Researchers must be explicit with the data analysis processes employed, ensuring analyses are clearly tied to research questions. Per \citet{braun2006using}, the interpretative power afforded in using thematic analysis is limited to description if not applied within a theoretical framework that grounds analytic claims.

Grounded theory is a widely used methodological approach, based on induction and comparison, used in qualitative investigation to construct theory \cite{glaser1967discovery, bryant2007sage}. Because the goal of grounded theory is theory development, researchers must ensure to not claim utilization of grounded theory when only using methods of coding regarded as part of---but not exclusive to---the coding canon of grounded theory: \emph{in vivo} coding, process coding, initial (open) coding, focused coding, axial coding, and/or theoretical (selective) coding \cite{Saldana2013}. In contrast to experimental research, grounded theory begins from empirical observations or data, not preformed theories and hypotheses \cite{LazarFengHochheiser2017}.

Indeed, promising avenues for qualitative research in XAI exist, particularly, with respect to interviewing. The Defense Advanced Research Projects Agency (DARPA) cites interviewing as a valuable method for studying mental models in XAI, namely, by combining tasks where users predict the decisions of an intelligent system with post-experimental interviews  \cite{MuellerHoffmanClanceyEmreyKlein2019}. Structured interviewing, in the form of a retrospection task with question-answering, is another example of a method that could prove useful for mental model elicitation in XAI \cite{HoffmanMuellerKleinLitman2018}. Others have used interviewing as part of research aimed at improving user trust in intelligible, context-aware machine learning applications \cite{Lim2012}. A lack of user trust is a key impediment to the uptake of AI applications \cite{Miller2019}. When studying how satisfied users are with explanations in XAI, `trustworthiness' is a core attribute used for measurement \cite{HoffmanMuellerKleinLitman2018}. Thus, we encourage XAI researchers to consider interviewing as a means to further investigate user trust in XAI.

Rigor is but one dimension of quality in qualitative research. Many other aspects also represent quality in qualitative investigation not illustrated as part of this analysis ({\it c.f.}, \citeauthor{Tracy2010} \citeyear{Tracy2010}). As techniques abound for capturing qualitative data, many of which could not be covered in this paper ({\it e.g.}, the think aloud protocol, diaries, user-centered design process), we contend non-experts in qualitative research should not be encumbered with the additional burden of designing, conducting, and analyzing the results of qualitative investigations in XAI. Ensuring rigor in the user studies conducted in XAI entailing qualitative investigation will promote cross-pollination with social experts outside the XAI community.

Echoing our sentiment from Section 1: we underscore the standpoints of \citet{PayrovnaziriChenRengifoMorenoMillerBianChenLiuHe2020}, \citet{BhattAndrusWellerXiang2020}, and \citet{Xu2019} and call for the XAI community to collaborate with experts from social disciplines toward bolstering rigor and effectiveness in user studies. The references in this paper afford a foundational basis for XAI researchers to become oriented with qualitative research and facilitate constructive dialogue with experts from social disciplines in support of collaboration. The XAI community is primed for a multidisciplinary convergence on user studies toward the benefit of increasing the trust users place in XAI systems. We hope this analysis will motivate some of the efforts required to converge on such multidisciplinary collaboration.

\subsection{Acknowledgments}
We would like to thank the chairs, committee members, and reviewers of the AAAI 2021 Explainable Agency in Artificial Intelligence Workshop. Adam Johs and Rosina Weber were supported by the National Center for Advancing Translational Sciences (NCATS), National Institutes of Health, through the Biomedical Data Translator program, award {\#}OT2TR003448. Any opinions expressed in this document are those of the authors and do not necessarily reflect the views of NCATS, other Translator team members, or affiliated organizations and institutions. We are also grateful to Prateek Goel for his assistance with formatting.

%%%%%END OF DOCUMENT%%%%%
%\bibliographystyle{aaai}
\bibliography{adam.bib} 
\end{document}